\documentclass[iop,apj,numberedappendix,twocolappendix,revtex4]{emulateapj}

\usepackage{epstopdf}

\usepackage{multirow}

\slugcomment{}

\shorttitle{Self-Gravitating Magnetized NDAF}
\shortauthors{Shahamat {\em et al.}}
\usepackage{graphicx}	% Including figure files
\usepackage{amsmath}	% Advanced maths commands
\usepackage{amssymb}	% Extra maths symbols
\usepackage{epstopdf}

\begin{document}

\title{Self-Gravity in Magnetized Neutrino-Dominated Accretion Discs}

\author{Narjes Shahamat $^{1}$ and Shahram Abbassi $^{1, 2}$}
\affil{$^1$Department of Physics, School of Sciences, Ferdowsi University of Mashhad, Mashhad, 91775-1436, Iran; abbassi@um.ac.ir}
\affil{$^2$School of Astronomy, Institute for Research in Fundamental Sciences (IPM), Tehran, 19395-5531, Iran}

\email{abbassi@um.ac.ir}

\begin{abstract}
 What is conducted in the present work is the study of self-gravity effects on vertical structure of magnetized neutrino-dominated accretion disc (termed NDAF and considered as a central engine for Gamma-ray bursts (GRBs)). Some of the disc physical time scales such as viscous, cooling and diffusion time scales, that are supposed to play a pivotal role in the late time evolutions of the disc, have been studied. It is of our interest to investigate the possibility of the X-ray flares occurrence, observed in late time GRB's extended emission through the "magnetic barrier" and "fragmentation" processes in our model. The results lead us to interpret self-gravity as an amplifier for Blandford-Payne luminosity (BP power) and the generated magnetic field, but a suppressor for neutrino luminosity and magnetic barrier process via highlighting fragmentation mechanism in the outer disc, especially for the higher mass accretion rates. 
\end{abstract}

% http://journals.aas.org/authors/keywords2013.html
\keywords{gamma-ray burst: general --- accretion discs --- neutrino --- magnetic field --- self-gravity}

\section{INTRODUCTION}
\label{introduction}

Gamma-ray bursts (GRBs) are sudden release of about $10^{51-54}erg$ energy in a volume with radius of less than $100Km$, which last from $0.01s$ till $100s$ (for reviews, see \citet{Piran2004}; \citet{Meszaros2006}; \citet{Nakar2007}; \citet{Gehrels et al.2009}; \citet{Kumar et al.2015}). According to the duration time $T_{90}$, that is defined as the time interval over which $90 \%$ of the total background-subtracted counts are observed, GRBs are usually classified into two classes: Long GRBs (LGRBs) ($T_{90}>2s$) whose existence emanate from the core collapse of massive stars (\citet{Kouveliotou et al.1993}; \citet{Woosley1993}; \citet{Paczynski1998}; \citet{Hjorth et al.2003}) and Short GRBs (SGRBs) ($T_{90}<2s$) whose origins are thought to be the coalescence of Neutron Stars (NS) or NS-black hole binary systems (\citet{Eichler et al.1989}; \citet{Narayan et al.1992}; \citet{Kouveliotou et al.1993}; \citet{Fryer et al.1998}). All these scenarios result in a hyperaccreting spinning stellar mass black hole with a mass accretion rate in the range of $0.01-10\frac{M_{\odot}}{s}$, surrounded by an accretion disc of several solar-mass (\citet{Popham et al.1999}; \citet{Gu et al.2006}; \citet{Liu et al.2007}), which is hot and dense enough to be cooled via neutrino pair annihilation on the surface (NDAF), or a magnetar (\citet{Usov1992}; \citet{Dai et al.2006};\citet{Yu et al.2013}). In addition to the NDAF model, the Blandford-Znajek (BZ) mechanism (\citet{Blandford et al.1977}) is considered as another promising candidate to cool this hyper-accretion disc, effectively (\citet{Meszaros et al.1997b}). 

In spite of these proposed mechanisms, the nature of GRB's central engine is still of great ambiguity. Considering the neutrino opacity, the neutrino-anti neutrino annihilation is not efficient enough to power the energetic short-hard GRBs (e.g., GRB080913). NDAF model is also unable to fuel the X-ray flares observed in GRB's emission by $Swift$ (\citet{Fan et al.2005}). On the other hand, Narayan, Paczynski and Piran (\citet{Narayan et al.1992}) showed that the magnetic field can reach $10^{12}G$ immediately after the disc formation and increases up to $10^{15-16}G$ due to shearing action of the differentially rotating disc. Hence, the idea of MHD process strengthened in later works (e.g., \citet{Di Matteo et al.2002}; \citet{Fan et al.2005};  \citet{Shibata et al.2007}). Furthermore, such a process might be a reliable alternative for X-ray flares through a magnetic barrier as discussed by Proga and Zhang (\citet{Proga et al.2006}). Accordingly, it seems rather inevitable to regard the effects of magnetic field in NDAF model.              

On the other hand, the high mass density of NDAFs ($\rho\sim 10^{10} g/ cm^{-3}$) motivates us to study the influence of self-gravity in the dynamical structure of these systems. With the idea that flares have something to do with what is in common between SGRBs and LGRBs (i.e., hyperaccretion disc), \citet{Perna et al.2006} argued that the gravitational instability leads either to large-amplitude changes in the inner accretion rate or complete fragmentation of the disc followed by a relatively slow in-spiral of fragments toward the black hole. This results in a less powered jet which can be the seed for the late time flares. Thus, taking the effects of self-gravity into account sounds to give us a more realistic insight into the NDAFs' structure and their evolution.     

 Some authors have studied the vertical structure of the optically thick disc, including NDAF (e.g., \citet{Gu et al.2007}; \citet{Liu et al.2010a}; \citet{Gu et al.2009}; \citet{Jiao et al.2009}). The impacts of vertical self-gravity has also been considered by Liu et al. (\citet{Liu et al.2014}). From the magnetic points of view, the study of NDAF structure has been developed through several works (e.g., \citet{Lei et al.2009}; \citet{Xie et al.2009}; \citet{Cao et al.2014}). However, combining these two significant physical features has remained to be concerned. In this paper, motivated by the above ideas, we have focused on the influence of vertical self-gravity in magnetized NDAF.
  
  In section~\ref{2.1}, the basic equations and assumptions are included. In order to consider the effects of self-gravity and study "magnetic barrier" and "fragmentation" probabilities, we introduce Toomre parameter and viscous, diffusion and cooling time scales in sections~\ref{2.2} and \ref{2.3}. Besides, neutrino and BP luminosities are included in section~\ref{2.4}. Section~\ref{3} represents our numerical outcomes, and the final discussion, containing the main conclusions, is given in section~\ref{4}.

\section{PHYSICAL MODEL}

\subsection{Basic Formalism}
\label{2.1}

We study a steady and axisymmetric magnetized NDAF ($\partial/\partial t=0$, $\partial/\partial \phi=0$) in which self-gravity has been taken into account, vertically. Considering the magnetic field influence in both large scale (magnetic braking mechanism) and small scale (viscous dissipation effects), through which the disc's rotational energy extraction and angular momentum transfer happen (\citet{Blandford1976}; \citet{Blandford et al.1982}; \citet{Balbus et al.1991}; \citet{Lee et al.2000}), yields the following results for the continuity and angular momentum equations     
\begin{equation}
\dot{M}=-2\pi R\varSigma v_{R}=constant
\label{(1)}
\end{equation}
\begin{equation}
\dot{M}=\frac{2\pi\alpha R^{2}\varPi}{\Omega_{k}R^{2}-j} + \frac{B_{\phi}B_{z}R^{2}} {\frac{\partial}{\partial R}{ (R^{2} \Omega_{k})}}
\label{(2)}
\end{equation}
where we have followed \citet{Lee et al.2000} and  \citet{Xie et al.2009} approach in order to magnetic braking effects being included. Note that $\dot{M}$ is the mass accretion rate, $v_{R}$ is the radial velocity, $\Omega_{k}=(GM/R)^{1/2}/(R-R_{g})$ is the Keplerian angular velocity, $G$ is the gravitational constant, $R_{g}=2GM/c^{2}$ is the Schwarzschild radius, $j=1.8cR_{g}$ is an integration constant representing the angular momentum of the innermost stable circular orbit (ISCO), $\alpha$ is the magnetic viscosity parameter (\citet{Shakura et al.1973}; \citet{Pringle1981}) and $B_{R}$, $B_{\phi}$ and $B_{z}$ are the three components of the magnetic field. Furthermore, $\varSigma$ and $\varPi$ are the surface density and vertically integrated pressure defined as 
\begin{equation}
\varSigma=2\int_{0}^{H}\rho dz
\end{equation}
\begin{equation}
\varPi=2\int_{0}^{H}p dz
\end{equation}
in which $p$ and $\rho$ are the disc pressure and mass density and the speed of sound is further defined as $c_{s}=(\frac{\varPi}{\varSigma})^{1/2}\approx(\frac{p_{0}}{\rho_{0}})^{1/2}$, where the zero index is related to the equatorial plane quantities. The half thickness of the disc is denoted by $H=\frac{\varSigma}{2\rho_{0}}$, here.  

The energy balance equation consists of viscous heating, Neutrino and advective cooling and the fraction of rotational energy extracted by magnetic braking effects (\citet{Xie et al.2009})
 \begin{equation}
Q_{vis}=Q_{adv}+Q_{\nu}^{-}+Q_{B}^{-}
\label{(5)}
\end{equation}
The viscous heating rate is
\begin{equation}
Q_{vis}=\frac{1}{2\pi}\dot{M}\Omega_{K}^{2}fg
\end{equation}
where $f=1-j/\Omega_{K}R^{2}$, and $g=-d\ln\Omega_{K}/d\ln R$. The advective cooling rate is
\begin{equation}
Q_{adv}=\frac{1}{2\pi}\frac{\xi\dot{M}c_{s}^{2}}{R^{2}} 
\end{equation}
in which $\xi=3/2$ is a dimensionless quantity of the order of unity (e.g., \citet{Kato et al.2008}). The neutrino cooling rate is expressed by a bridging formula (e.g., \citet{Di Matteo et al.2002}; \citet{Kohri et al.2005}) as follows:
\begin{equation}
Q_{\nu}=\underset{i}{\sum}\frac{(7/8)\sigma T^{4}}{(3/4)[\frac{\tau_{\nu_{i}}}{2}+\frac{1}{\sqrt{3}}+\frac{1}{3\tau_{a,\nu_{i}}}]}
\end{equation}
where $\sigma$ is the Stefan-Boltzmann constant, $T$ is the temperature, and $\tau_{\nu_{i}}$ is the total optical depth for neutrinos, including the absorption optical depth $\tau_{a,\nu_{i}}$ and scattering optical depth $\tau_{s,\nu_{i}}$ 
\begin{equation}
\tau_{\nu_{i}}=\tau_{a,\nu_{i}}+\tau_{s,\nu_{i}}
\end{equation}
where $i={1,2,3}$ refers to the three kinds of leptons, $\nu_{e}$, $\nu_{\mu}$ , and $\nu_{\tau}$.  The main absorption processes include the electron-positron pair annihilation and Urca processes (e.g., \citet{Narayan et al.2001}; \citet{Di Matteo et al.2002}; \citet{Liu et al.2014}), and the corresponding optical depths can be written as
\begin{equation}
\tau_{a,\nu_{i}}=2.5\times 10^{-7}T_{11}^{5}H
\end{equation}
\begin{equation}
\tau_{a,\nu_{e}}=2.5\times 10^{-7}T_{11}^{2}X_{nuc}\rho_{10}H
\end{equation}
where $T_{11}=\frac{T}{10^{11}}K$ , $\rho_{10}=\frac{\rho}{10^{10}}~ gcm^{-3}$ , and $X_{nuc}$ is the mass fraction of free nucleons approximately given by (e.g., \citet{Liu et al.2014})
\begin{equation}
X_{nuc}=min\{1,295.5\rho_{10}^{-3/4}T_{11}^{9/8}exp(-0.8209/T_{11})\}
\end{equation}

The scattering optical depth by nucleons can be given by
\begin{equation}
\tau_{s,\nu_{i}}=2.7\times 10^{-7}T_{11}^{2}\rho_{10}H
\end{equation}
Fallowing \citet{Lee et al.2000} and \citet{Xie et al.2009}, $Q_{B}^{-}$ can be calculated as $\varOmega d\tau/ds$, in which $d\tau$ is the torque exerted by the annular ring with width $dR$ of the disc due to the Lorentz force ($ds=2\pi RdR$ ). Thus, we have
\begin{equation}
Q_{B}^{-}=2R\varOmega(B_{\phi}B_{z}/4\pi)
\label{(14)}
\end{equation}

The equation of state (EOS) is written as (e.g., \citet{Di Matteo et al.2002}; \citet{Liu et al.2014})
\begin{equation}
p=p_{gas}+p_{rad}+p_{deg}+p_{\nu}
\label{(15)}
\end{equation}
The gas pressure from the free nucleons and $\alpha$-particles is $p_{gas}=\frac{\rho k_{B}T}{m_{p}}\frac{1+3X_{nuc}}{4}$, where $a$ is the radiation constant, $T$  is the disc temperature and $k_{B}$ is the Stephan-Boltzman constant (see \citet{Woosley et al.1992}; \citet{Qian et al.1996}). The radiation pressure is $p_{rad}=\frac{1}{3}aT^{4}$. In degeneracy term $p_{deg}=\frac{2\pi hc}{3}(\frac{3\rho}{16\pi m_{u}})^{4/3}$, where $m_{u}$ is the mean mass of a nucleon and $h$ is the Planck constant, only the degeneracy of electrons (rather than both electrons and nucleons) has been regarded. Generally speaking, the degeneracy pressure is important at high density and low temperature regimes. Such regimes inevitably appear in very massive discs, like NDAFs. Last term denotes the neutrino pressure $P_{\nu}=\frac{u_{\nu}}{3}$, in which $u_{\nu}=\underset{i}{\sum}\frac{(7/8)aT^{4}(\tau_{\nu_{i}}/2+1/\sqrt{3})}{\tau_{\nu_{i}}/2+1/\sqrt{3}+1/(3\tau_{a,\nu_{i}})}$ is the neutrino-energy density (\citet{Di Matteo et al.2002}; \citet{Kohri et al.2005}). Giving the correct behavior in the limit of both small and large $\tau_{\nu_{i}}$ and $\tau_{a,\nu_{i}}$, this relation was derived in the context of radiative transport and assumes that opacities and emissivities are independent of the $z$ coordinate. However, this assumption might be less accurate for the neutrino transport where all cross sections are a function of temperature and hence of the vertical disc structure.

Moreover, we consider the polytropic equation of state in vertical direction $p=K\rho^{4/3}$, where $K$ is a constant (\citet{Liu et al.2014}).

Another balance equation of which we are in need is the hydrostatic balance equation in vertical direction that reads
\begin{equation}
 4\pi G\varSigma_{z}+\frac{\partial\Psi}{\partial z}+\frac{1}{\rho}\frac{\partial p}{\partial z}+\frac{1}{8\pi\rho}(\frac{\partial B_{\varphi}^{2}}{\partial z}+\frac{\partial B_{R}^{2}}{\partial z})-\frac{1}{4\pi\rho}B_{R}\frac{\partial B_{z}}{\partial R}=0
 \label{(16)}
\end{equation}
In the case of $\varSigma_{z}=\int_{0}^{z}\rho dz^{\prime}$ (\citet{Paczynski1978a},\citet{1978b}) being slowly varied with radius, the first term represents the vertical self-gravity. The two last terms are the $\phi$-component of Lorentz force, as well.
Now we adopt the pseudo-Newtonian potential, written by Paczynski and Wiita (\citet{Paczynski et al.1980}), in order to mimic the effective potential  of a Schwarzschild black hole:
\begin{equation}
\Psi=\frac{-GM}{\sqrt{R^{2}+z^{2}}-R_{g}}
\label{(17)}  
\end{equation}

The consideration of Gauss's law, $\nabla.B=0$, the polytropic equation of state and equation (\ref{(17)}) leads to

\begin{eqnarray}
 && 4\pi G\varSigma_{z}+\Omega^{2}z+\frac{4}{3\rho^{2/3}}k\frac{\partial \rho}{\partial z}-\frac{1}{4\pi\rho}[(\frac{\partial (B_{R}B_{z})}{\partial R}+B_{z}\frac{\partial B_{z}}{\partial z})\nonumber\\\nonumber\\
 && +\frac{B_{z}B_{R}}{R}-\frac{1}{2}\frac{\partial }{\partial z}(B^{2}-B_{z}^{2}+B_{R}^{2}+B_{\phi}^{2})]=0 
 \label{(18)}
 \end{eqnarray} 
 
 As the next step, we make use of the local shearing sheet simulations of accretion discs in which it was found that the magnetic field components have somewhat standard ratios (\citet{Xie et al.2009}). For example, table 2 of \citet{Stone et al.1996} gives (note that their $x, y, z$ correspond to $R, \phi, z$)
  \begin{equation}
 B_{R}^{2}\simeq \frac{1}{10} B_{\phi}^{2}
 \label{(19)}
 \end{equation}
 \begin{equation}
 B_{z}^{2}\simeq \frac{1}{20} B_{\phi}^{2}
 \label{(20)}
 \end{equation}
and
 \begin{equation}
 B_{z}B_{R}\simeq4\pi\times10^{-5}p
 \label{(21)}
  \end{equation} 
  
We also need to have some additional considerations, essential for the hydrostatic balance equation to be vertically integrated, as follows: mathematically speaking, the magnetic pressure, $p_{mag}=B^{2}/8\pi$, can be written as
  \begin{equation}
 B^{2}/8\pi<p+B^{2}/8\pi\nonumber\\
  \end{equation}
 which yields
  \begin{equation}
  p_{mag}=\beta(p+B^{2}/8\pi)\ with\ \beta<1.\nonumber\\
  \end{equation}
  Therefore
  \begin{equation}
  B_{0}^{2}=8\pi\frac{\beta_{1}}{1-\beta_{1}}p_{0}
  \end{equation}
 and
 \begin{equation}
 B_{H}^{2}=8\pi\frac{\beta_{2}}{1-\beta_{2}}p_{H}.
 \end{equation}
 On the other hand
 \begin{equation}
 B_{z0}^{2}<B_{0}^{2}\nonumber\\
 \end{equation}
 so
 \begin{equation}
  B_{z0}^{2}=\gamma_{1}B_{0}^{2} ~with~ \gamma_{1}<1
 \end{equation}
 and similarly
 \begin{equation}
  B_{zH}^{2}=\gamma_{2}B_{H}^{2} ~with~ \gamma_{2}<1
 \end{equation} 

As the last step, we use Taylor expansion for density about $"z=0"$ up to second order term (as a good approximation for thin accretion discs) besides the reflection symmetry for density, and consider $\rho_{H}=\rho_{0}exp(-1/2)$ (\citet{Cao et al.2014}) ($"H"$ index refers to the surface quantities). Thus we have
 \begin{equation}
 \rho=\rho_{0}+\alpha_{1}\frac{z^{2}}{2H^{2}}\nonumber
 \end{equation}
 where
 \begin{equation}
 \alpha_{1}=2\rho_{0}(e^{-1/2}-1)\nonumber
 \end{equation}
 
 Regarding all above considerations, in addition to $p_{H}=p_{0} exp(-1/2)$, one can integrate equation (\ref{(18)}) over $"z"$, which yields
\begin{eqnarray}
&& 2\pi G\rho_{0}^{2}H^{2}+\Omega^{2}\frac{\rho_{0}H^{2}}{2}+p_{0}(e^{-1/2}-1)
-10^{-5}H\frac{dp_{0}}{dR}\nonumber\\\nonumber\\
&&-2 p_{0}(\frac{\beta_{2}}{1-\beta_{2}}\gamma_{2}e^{-2/3}+\frac{\beta_{1}}{1-\beta_{1}}\gamma_{1})-\frac{10^{-5}p_{0}H}{R}\nonumber\\\nonumber\\
&&+ p_{0}(\frac{\beta_{2}}{1-\beta_{2}}e^{-2/3}+\frac{\beta_{1}}{1-\beta_{1}})=0
\label{(26)}
\end{eqnarray}
For the sake of simplicity, we ignore the zero indices.

Now we are left with eight unknowns ($v_{R}, \rho, p, h, T$ and three components of magnetic field) and seven equations ~(i.e., (\ref{(1)}),(\ref{(2)}),(\ref{(5)}),(\ref{(15)}),(\ref{(19)}),(\ref{(20)}) and (\ref{(26)})). Thus we are in need of one more equation and a boundary condition for the disc pressure, since equation (\ref{(26)}) is a differential one. The latter requirement can be disappeared by applying the numerical results obtained by some authors such as \citet{Popham et al.1999}, which estimates a value of about $10^{30} erg/cm^{3}$ for the pressure in the inner regions of such hyperaccretion discs. Furthermore, the magnetic viscosity equation, as the last required equation, seems to be worthwhile here
 \begin{equation}
\frac{B_{R} B_{\phi}} {4\pi} =-\frac{3}{2} \alpha p.
\label{(27)}
\end{equation}

In order to achieve a better understanding of the effects of magnetic field and self-gravity through an analogy among the three cases: self-gravitating magnetized NDAF, magnetized case and self-gravitating one, we may get the required equations by ignoring the terms associated with the magnetic field and self-gravity in the above mentioned equations. For instance, considering the hydrostatic balance equation (\ref{(16)}), the magnetized NDAF can be obtained via ignoring the first term (which reflects the self-gravity impact)
 \begin{equation}
  \frac{\partial\Psi}{\partial z}+\frac{1}{\rho}\frac{\partial p}{\partial z}+\frac{1}{8\pi\rho}(\frac{\partial B_{\varphi}^{2}}{\partial z}+\frac{\partial B_{R}^{2}}{\partial z})-\frac{1}{4\pi\rho}B_{R}\frac{\partial B_{z}}{\partial R}=0.
  \label{(28)}
\end{equation}  
The self-gravitating case is also achievable through an elimination of the magnetic terms
\begin{equation}
 4\pi G\varSigma_{z}+\frac{\partial\Psi}{\partial z}+\frac{1}{\rho}\frac{\partial p}{\partial z}=0.
 \label{(29)}
\end{equation} 
After regarding the similar considerations to get the equation (\ref{(26)}), it can be written in the form of  
\begin{eqnarray}
&& 2\pi G\rho_{0}^{2}H^{2}+\Omega^{2}\frac{\rho_{0}H^{2}}{2}+p_{0}(e^{-1/2}-1)=0.
\label{(30)}
\end{eqnarray}

\subsection{Toomre Parameter}
\label{2.2}

There are several numerical theoretical studies and simulations that describe a massive unstable discin Newtonian gravity (e.g., \citet{Bonnell1994}; \citet{Matsumoto et al.2003}) or even in Modified Gravity (e.g., \citet{Roshan et al.2015}; \citet{Roshan et al.2016}). However, considering existing numerical simulations, it is not possible simply to say that all massive discs fragment because, to our knowledge, not only do current numerical simulations suffer from their own limitations, but they also do not generally give a fully consistent picture of global fragmentation. This situation gets more complicated when we consider magnetic fields. Apparently, the relation between gravitational instability and MHD turbulence is rather sophisticated, yet, the more the MHD turbulence is getting highlighted the less the gravitational instability comes to notice (\citet{Fromang2005}, \citet{Shadmehri et al.2006}). Moreover, the fact that the accretion discs are differentially rotating is against gravitational collapse. Anyway, it is the Toomre parameter that determines whether our NDAF is gravitationally unstable. In non-magnetic literature, this criterion reads 
\begin{equation}
Q=\frac{c_{s}\Omega_{k}}{\pi G \Sigma_{z}}
\end{equation}
    where $Q<1$ implies instability. However, in the presence of magnetic field, this will be modified by a factor of $(1+\frac{\beta}{1-\beta})$ as follows (e.g., \citet{Shu1992})
     \begin{equation}
Q_{M}=Q(1+\frac{\beta}{1-\beta})
\end{equation}
where $\beta$ has been introduced in section ~\ref{2.1}.

\subsection{"Magnetic Barrier" And/Or "Fragmentation"?}
\label{2.3}

 Proposed by \citet{Proga et al.2006}, the mass accretion rate's decline during the late time evolution of a hyperaccretion system can be considered as the source of unexpected X-ray flares in GRBs. They argued that the accumulated magnetic flux, in the inner edge of the disc, is capable of halting the accretion flow, intermittently. The importance of the magnetic effects for the X-ray flares can also be considered based on the energy budget of the accretion model (\citet{Fan et al.2005}). Following \citet{Xie et al.2009}, we take the advantage of the two timescales comparison, namely diffusion and viscouse timescales,  in order to investigate the possibility of this process. 
 
 The magnetic field buoyancy and its rising time toward the disc surface can be estimated as
   \begin{equation}
  t_{dif}\approx \frac{H}{v_{A}}
  \end{equation}
  where $v_{A}$ is the Alfven velocity, $v_{A} = \frac{B}{(4\pi\rho)^{1/2}}$. Besides, since the field lines are frozen to the disc materials, the viscous time $t_{\nu}$ can present the timescale of the magnetic flux accumulating in the vicinity of the black-hole, then we have
   \begin{equation}
    t_{\nu}=\int_{3R_{g}}^{R} \frac{1}{v_{R}} dR
    \end{equation}
    
  In the gravitational context, \citet{Perna et al.2006} discussed that in the case of instabilities either a quasi-spiral structure may be imposed onto the disc by which the large-amplitude outbursts of accretion can be derived if the disc mass is sufficiently large (\citet{Laughlin et al.1998}; \citet{Lodato et al.2005}), or the disc may fragment into bound objects. The latter is inevitable if (\citet{Gammie2001}; \citet{Perna et al.2006})
    \begin{equation}
        t_{cool}<t_{cirt}\approx 3\Omega^{-1},
        \end{equation}
        where cooling timescale is denoted by $t_{cool}\approx(H/R)^{2}t_{\nu}$ (\citet{Pringle1981}).  Through such an analogy, the possibility of fragmentation can be verified in our model. 
  
  \subsection{Neutrino Luminosity}
   \label{2.4}
          After neutrino cooling rate $Q_{\nu}$ being calculated, we are able to measure the neutrino luminosity, $L_{\nu}$, that is expressed as
           \begin{equation}
          L_{\nu} =2\pi\int_{R_{in}}^{R_{out}}Q_{\nu}RdR
          \end{equation}
          where we adopt $R_{in}=3R_{g}$ and $R_{out}=200R_{g}$.
          
          In the presence of strong magnetic field, it is also possible to extract the rotational energy of the disc or rotating black hole through ultrarelativistic GRB outflows driven as Pointing flux. Blandford-Znajek (BZ) and Blandford-Payne (BP) processes are the two proposed mechanisms through which this purpose is fulfilled. The first suggests that the black hole rotational energy can be extracted by large-scale magnetic fields threading the horizon (\citet{Blandford et al.1977}). The latter asserts that an outflow of matter can be driven centrifugally by large-scale magnetic fields anchored at the surface of the disc (\citet{Blandford et al.1982}).  Hence, for the BP process the risk of baryonic pollution is much larger as the wind originates from high-density regions (\citet{Daigne et al.2002}). These winds of different Lorentz factors launched by BZ and BP processes compose a spine/sheath jet structure, which was first pointed out by Meier (2003) from observational reasons and presented by Wang et al. (2008) to explain the jets for active galactic nuclei and black hole binaries. Nonetheless, following \citet{Xie et al.2009}, we consider the luminosity provided by this process since our central black hole is of the form of Schwarzschild black hole not a spinning one. The BP power output from a disc is equal to the power of disc magnetic braking and can be calculated as (\citet{Livio et al.1999}; \citet{Lee et al.2000})
           \begin{equation}
                  L_{BP} =2\pi\int_{R_{in}}^{R_{out}}Q_{B}RdR.
                  \end{equation}

        \begin{figure*}
                            \centering
                            \includegraphics[scale=0.81]{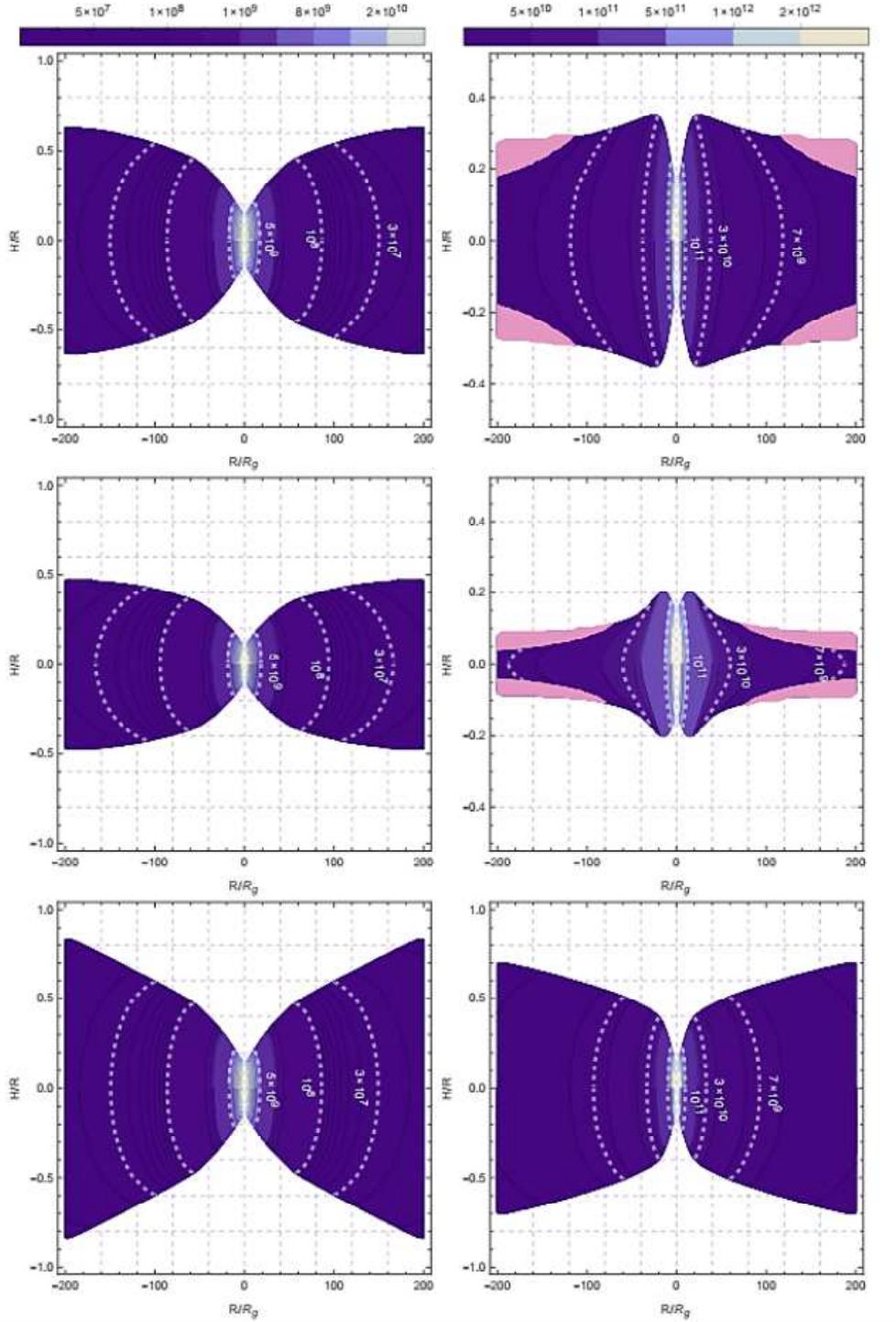}

                                \caption{Contours of density for $\dot{M}=0.1M_{\odot}/s$ (three left plots) and $\dot{M}=10M_{\odot}/s$ (three right plots), with three orders of density are highlighted through dashed lines ($5\times 10^{9},~10^{8},~3\times~10^{7}~gr/cm^{3}$ for the left panels, and $10^{11},~3\times 10^{10},~7\times~10^{9}~gr/cm^{3}$ for the right ones ). The two top panels are devoted to the self-gravitating magnetized NDAF, the tow middle plots are related to the self-gravitating case and the two bottom ones show the magnetized NDAF. Additionally, the shaded areas in pink show the gravitationally unstable zones.   }
                                \label{fig1}
                            \end{figure*}

  \section{NUMERICAL RESULTS} 
                                                                                                             \label{3}
                                                                                                             As it was pointed out in section~\ref{2.1}, our set of eight equations ~((\ref{(1)}),(\ref{(2)}),(\ref{(5)}),(\ref{(15)}),(\ref{(19)}),(\ref{(20)}),(\ref{(26)}) and (\ref{(27)})) can be solved numerically, in which we have fixed $\alpha = 0.1$ and $M =3M_{\odot}$. In spite of the inferred interval for constants $\beta_{1},\beta_{2},\gamma_{1}$ and $\gamma_{2}$, we estimate their values based on simulations such as \citet{Stone et al.1996}, from which one can deduce $\beta_{1}\simeq\beta_{2}\simeq 0.01$ and $\gamma_{1}\simeq \gamma_{2}\simeq 0.04$.
                                                                                                              
                                                                                                               In Figure~\ref{fig1}, for two different mass accretion rates ($\dot{M}=0.1M_{\odot}/s$ (the three left panels) and $\dot{M}=10 M_{\odot}/s$ (the three right panels), we have outlined the contours of density through the whole disc with the gravitationally unstable regions ($Q<1$) have been shaded in pink. We have considered self-gravitating magnetized NDAF in the top two plots, self-gravitating case in the middle two panels, and the two last ones are the magnetized case. 
                                                                                                               
                                                                                                               First of all, one can see self-gravity has made the disc getting thinner, especially for the higher accretion rates and in the outer regions, where self-gravity play an important role. It is also worth noting that the magnetic field has opposed the self-gravity and caused the disc scale-hight to increase.  
                                               Secondly, the density drops by around three orders of magnitude when we go outward, radially. In the case of self-gravitating magnetized NDAF, the density is getting larger compared to the magnetized NDAF. Of course, in agreement with \citet{Liu et al.2014}, such a behavior arises in the outer regions and it is getting more obvious as the accretion rate grows.  
                                                                                                                                                      Furthermore, one can trace the gravitational instabilities ($Q<1$) via the shaded realms in purple. Similar to \citet{Liu et al.2014}, as the mass accretion rate increases the unstable realms grow inwards, such that in the case of $\dot{M}=0.1M_{\odot}/s$ we are not encountered with instability, but it is not the case for $\dot{M}=10 M_{\odot}/s$. And last but not least, the magnetic field appears as a suppressor against gravitational instabilities, so that the unstable regions shrink when the magnetic effects come into play. This fully agrees with our expectations based on section~\ref{2.2}.
                                                                            To probe the possibility of  fragmentation, we can make estimations for self-gravitating magnetized NDAF based on the mentioned time scales in section~\ref{2.3}, i.e. cooling and critical time scales. In the outer disc, which is more likely to be gravitationally unstable, the cooling time scale is of the order of about $0.01 s$ for $\dot{M}=4 M_{\odot}/s$ (that is the start point at which instability occurs in the area of our interest), and about $0.001 s$ for $\dot{M}=10 M_{\odot}/s$. These are obviously less than the critical time scale that is of the order of $0.3 s$ in the outer zones. Thus, fragmentation process would be a probable mechanism to make the mass accretion rate descend and, consequently, leads to the late time X-ray flares, especially in the case of higher accretion rates.                                                                       
                                                                                                                                          
                            \begin{figure}
                                             \includegraphics[scale=0.8]{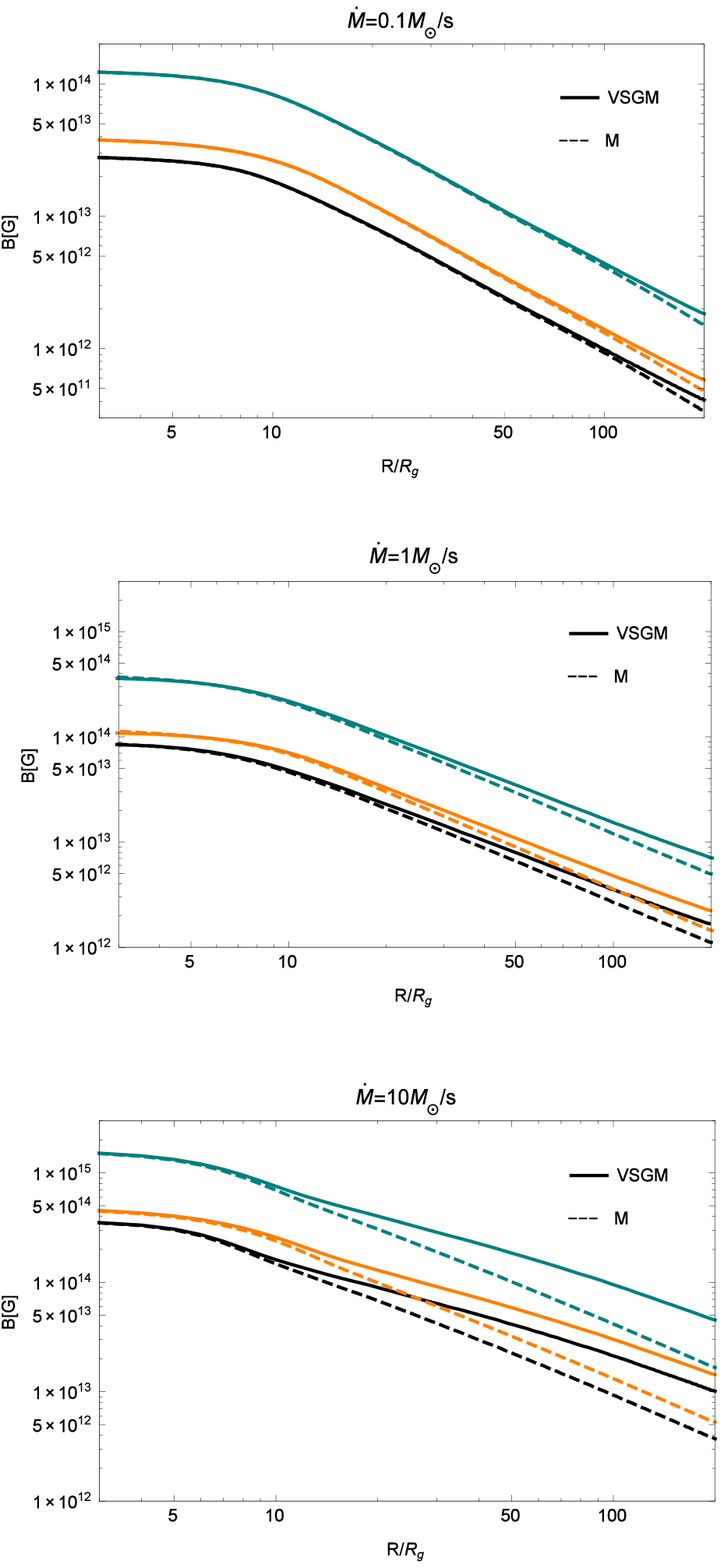}
                                             
                                                \caption{Magnetic field components, $B_{R},B_{\phi},B_{z}$, for two cases of magnetized NDAF: with (VSGM) and without self-gravity (M) (solid and dashed curves). In these three panels, the radial, azimuthal and poloidal components are plotted in orange, blue and black, respectively. }
                                                \label{fig2}
                                            \end{figure}

                                    \begin{figure}

                                                                             \includegraphics[scale=0.69]{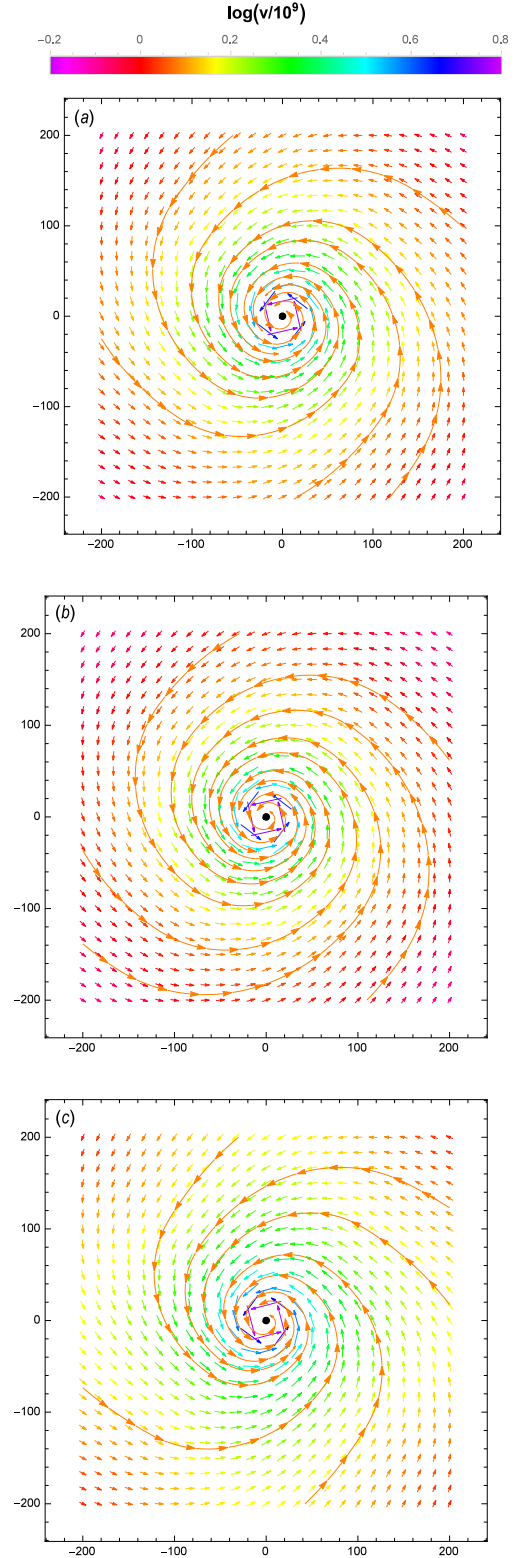}
                                                                             
                                              \caption{ The illustration of velocity vector field ($cm/s^{-1}$) on the disc's equatorial plane, for $\dot{M}=10 M_{\odot}/s$, with some stream lines have been outlined. In case (a) both self-gravity and magnetic field have been included, but it is the magnetic field that governs the disc in plot (b). And the last one, plot (c), only covers the self-gravity impacts.           }
                                                                 \label{fig3}                                  
                                                                    \end{figure}  
                                                                    
                                                                     \begin{figure}
                                                                                                                                                                                               \includegraphics[scale=0.8]{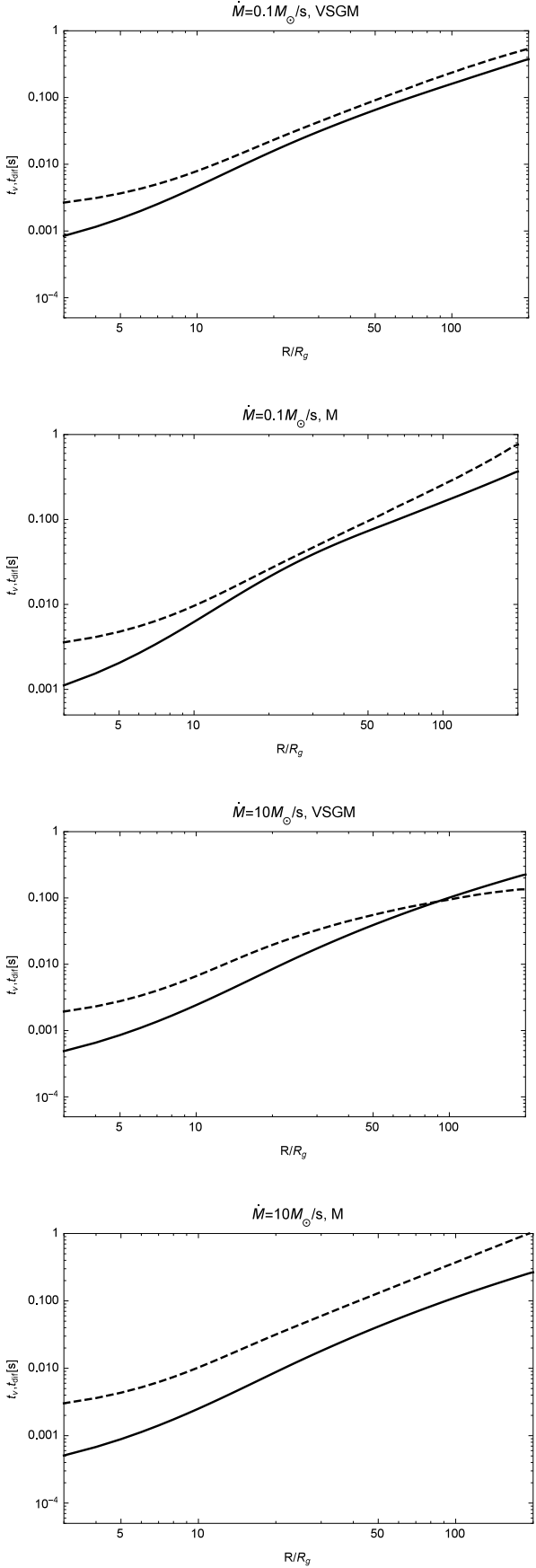}
 \caption{Viscous and diffusion time scales for $\dot{M}=0.1,10 M_{\odot}/s$ illustrated by solid and dashed lines. Both magnetized (M) and vertically self-gravitating magnetized (VSGM) cases are considered.     }
                                                                                                                                                                                                  \label{fig4}
                                                                                                                                                                                              \end{figure}
                                                                                                                                                         \begin{figure}
                                                                                                                                                                                                     \includegraphics[scale=0.8]{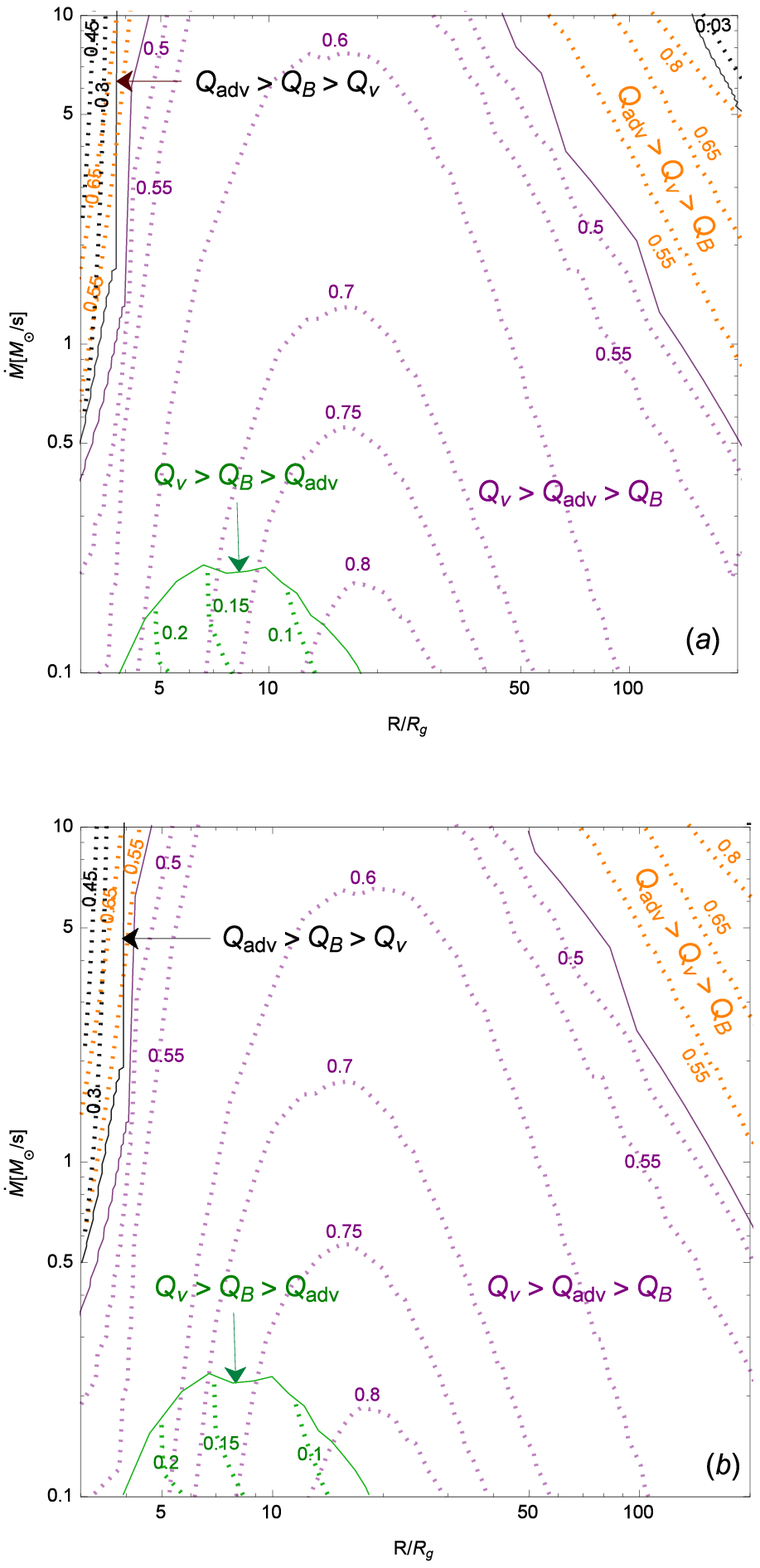}

                                                                                                                                           \caption{ The contours of three different cooling rates' ratios with respect to viscose cooling rate (i.e., $Q_{adv}/Q_{vis},~Q_{B}/Q_{vis}$ and $Q_{\nu}/Q_{vis}$), in $R-\dot{M}$ plane. The boundaries of their dominancy are shown with solid lines. Plot (a) is the case for self-gravitating magnetized NDAF, whereas plot (b) represents the magnetized one. }  
                                                                                                                                                                                                            \label{fig5}                           
                                                               \end{figure}

                                                                                                             The magnetic field components, $B_{R},B_{\phi},B_{z}$, for the three mass accretion rates ($\dot{M}=0.1,1,10M_{\odot}/s$), have been plotted in Figure~\ref{fig2}. Inwardly going through the disc, they show a rise of about two orders of magnitude, up to $10^{14-15}G$, which is strong enough to power the most energetic GRBs with the luminosity of about $10^{53}erg/s$. Furthermore, as self-gravity is taken into account the magnetic field increases, which might reflect the dependency of the seed magnetic field's generation on the vertical density profile of the disc. This has been proposed by \citet{Safarzadeh et al.2017} through a recent study, in which it is argued that the radial temperature profile and the vertical density profile of accretion discs provide the necessarily conditions for "Biermann battery" process (that is a responsible mechanism for the generation of the seed magnetic field) to operate naturally. 
                                                                                                             
                                                                                                             Figure~\ref{fig3} is an illustration of velocity vector field on the disc's equatorial plane, for $\dot{M}=10 M_{\odot}/s$, with some stream lines have been outlined. The change in colors expresses the variation of vectors' magnitude in logarithmic scale. The first panel (a) includes the magnetic effects as well as self-gravity. Comparing (a) with (b), in which self-gravity has been ignored, the velocity grows as an impact of self-gravity. On the other hand, plot (c) does not contain magnetic field and considers the vertical self-gravity, instead. Through an analogy between this panel and (a), one can find a reduction in velocity magnitude as the magnetic field comes into play.

                                                                                                          The probability of "magnetic barrier" occurrence can be inferred from Figure~\ref{fig4}, which provides us with the power of comparing the diffusion and viscous time scales for two mass accretion rates ($\dot{M}=0.1,10 M_{\odot}/s$), with and without self-gravity. First, in the lower accretion rates, the viscous time is sufficiently less than the diffusion time, that can enhance the chance of magnetic field accumulation and subsequently magnetic barrier. Next, in the absence of vertical self-gravity the diffusion time experiences an increase in the outer regions where the self-gravity appears to be more effective. Such an effect implies that the vertical diffusion or buoyancy process is facilitated by self-gravity since it makes the disc thinner. Yet, it seems to have no effect on viscous time scale which reveals the fact that we have just considered the $r\phi$-component of the viscous stress tensor. Moreover, as the mass accretion rate increases, which lead to a growth in density and a drop in the disc scale hight (Figure~\ref{fig1}), both diffusion and viscous time scales decline. Lastly, in the case of higher accretion rates, self-gravity may suppress the magnetic barrier.       
                                                                                                                   
\begin{figure}
                                                                                                                                   \includegraphics[scale=0.85]{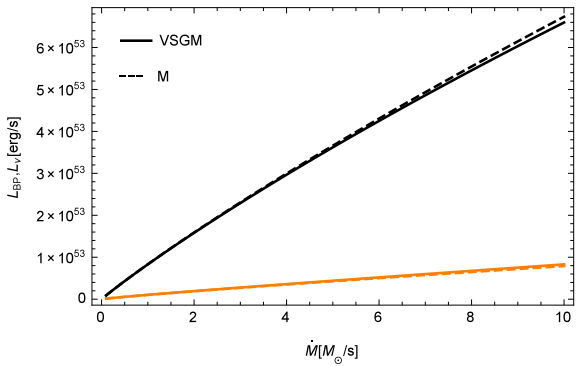}
                                                                                                                                  \caption{Neutrino luminosity and BP power (black and orange lines) versus mass accretion rate. Solid lines represent the magnetized NDAF with vertical self-gravity (VSGM), but in dashed lines it is ignored (M).}
                                                                                                                                      \label{fig6}
                                                                                                                                  \end{figure}

\begin{figure}
                                                                                                                                   \includegraphics[scale=0.8]{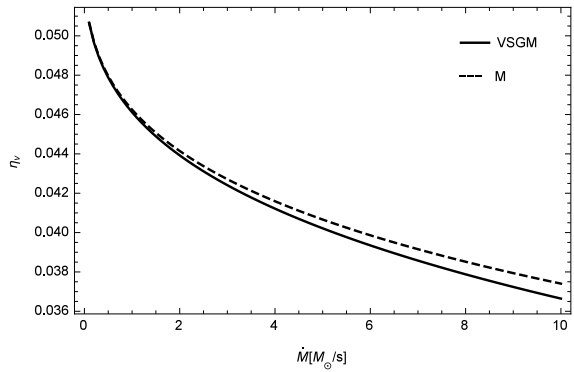}
                                                                                                                                  \caption{ Neutrino cooling efficiency versus mass accretion rate. Solid and dashed curves display the magnetized NDAF considering (VSGM) and ignoring self-gravity (M), respectively.  }
                                                                                                                                      \label{fig7}
                                                                                                                                  \end{figure}

                                                                                                          The zones where are dominated by each of the cooling rates (i.e., neutrino, advection and magnetic field fractions), in $R- \dot{M}$ plane, are presented in Figure~\ref{fig5}. Plot (a) considers both self-gravity and magnetic field and (b) is related to the magnetic effects. Also, it should be mentioned that the green and black boundaries, with their contours of $Q_{B}/Q_{vis}$ ratio, correspond to the magnetic cooling dominated realms over advective and neutrino processes, respectively. The contours in purple display neutrino dominated regions over all other cooling processes. And orange contours correspond to the advection dominated areas over two other cooling mechanisms. Generally speaking, neutrino cooling gets highlighted over an extended area. Moreover, advetive cooling rate becomes more effective as self-gravity is taken into account, specially in the outer disc which is more affected by self-gravity. Such an outcome might arise as a result of a growth in density and, subsequently, the generated magnetic field will grow. This causes the magnetic process gets more efficient in the outer disc, as well.
                                                                                           Figure~\ref{fig6} is an illustration of BP power and neutrino luminosity Vs. mass accretion rate throughout the disc. Both cases sound to be strong enough to power the GRB jets. On the other hand, the smooth downward trend of neutrino cooling efficiency (Figure~\ref{fig7}) ($\eta_{\nu}=L_{\nu}/\dot{M}c^{2}$, with which energy is transported out of the flow by neutrinos) can display the reduction of neutrinos capability to cool the disc. Such a decline is caused by the increase of neutrino opacity for higher accretion rates (\citet{Di Matteo et al.2002}). This leads neutrinos to be more trapped. Furthermore, self-gravity appears to play opposite roles in two cases (BP power and neutrino luminosity). That is to say, it has lessened the luminosity of neutrinos, which is in contrast to the \citet{Liu et al.2014} outcomes, but enhanced the BP power. This unexpected behavior reflects the fact that self-gravity affects BP power more than neutrino luminosity to be strengthened. For one thing, neutrino emission occurs more in the inner disc, but it is the outer regions that are more affected by self-gravity. For another, self-gravity increases the generated magnetic field and, consequently, highlights its role to cool the disc, which may weaken neutrinos' fraction. 
  
  Ignoring the nucleons' degeneracy, there might be some worth noting points. Firstly, \citet{Kohri et al.2002} have argued that the nucleon degeneracy, besides that of the electron, has a suppressing effect on the neutrino cooling rate and this subsequently lowers the neutrino luminosity and its efficiency. The second is how fragmentation and magnetic barrier would be affected by such a consideration. We think that, since the nature of degeneracy pressure is against the density enhancement due to the self-gravity, the possibility of fragmentation should drop if the nucleon degeneracy is taken into account. This might be the same for the magnetic barrier, in which the nucleon degeneracy will result in a raise in the disc's pressure and consequently the density will grow (consider the polytropic equation of state). Of course, this relation might be inferable from  \citet{Kohri et al.2002} that confirms the denser regions correspond to the realms with higher nucleon degeneracy importance. This can lead to a stronger magnetic field, as Figure~\ref{fig2} reflects such an impact. Thus, the magnetic barrier might get less probable due to a decrease in diffusion timescale and a growth in viscous timescale, each of which are a result of the stronger magnetic field.

                                                                   \section{Estimates And Observational Evidences} 
                \label{4}
                
 In this section, following the \citet{Proga et al.2006}, we are to make estimates of some X-ray flare's features. At first step, regarding our results, we calculate the magnetic flux accumulated in the vicinity of the black hole and test if it has the capability to result in the X-ray flares. To take the magnetic barrier into account, the consideration of which we are in need is that the radial magnetic force, $F_{m}\approx 2B_{R}B_{z}/4\pi$, should balance the gravitational one, $F_{g}=GM_{b}\Sigma/R^{2}$, to support the in falling gas. $\Sigma = \dot{M}/(2\pi R\epsilon v_{ff})$ is the surface density of the gas and $\epsilon v_{ff}$ is the flow radial velocity assumed to be a fraction $\epsilon$ of the free fall velocity, $v_{ff} =(2GM_{b}/R)^{1/2}$ (we refer readers to \citet{Narayan et al.2003} for more information about "magnetically arrested disc"). Assuming $B_{R}\approx B_{z}=B$, which is approximately consistent with the simulation results we have applied, the force balance yields the magnetic flux as $\Phi \approx \pi R^{2}B(r)=5\times 10^{28} \epsilon_{-3}^{-1/2}(R/R_{g})^{3/4}\dot{M}_{1}^{1/2}M_{3}~cm^{2}G$, where $\epsilon_{-3}=10^{3}\epsilon$, $\dot{M}_{1}=\dot{M}/1M_{\odot}~s^{-1}$ and $M_{3}=M_{b}/3M_{\odot}$. The magnetospheric radius, at which the accumulated poloidal field disrupts the accretion flow and it lies well outside the event horizon of the black hole, can be estimated as $r_{m}\approx 60\epsilon_{-3}^{2/3}\dot{M}_{1}^{-2/3}M_{3}^{-4/3}\Phi_{30}^{4/3}$, where $\Phi_{30} \equiv \Phi/(10^{30}~cm^{2}G)$. 
 
 To estimate the conditions needed to restart accretion, the accretion energetics and related time-scales, we ask what is the mass of a disc with $ \Sigma $ high enough to reduce $ r_{m} $ from its own magnitude in the late time evolution to 3 or so. Before that we should make an estimation for $r_{m}$ in the late time disc's activity. Although, there are somehow different strategies in the literature to make this estimate, e.g. \citet{Ronning et al.2016} and \citet{ Tchekhovskoy et al.2014}, we keep following the \citet{Proga et al.2006}. In the case that the hyper-rate of $ 10~M_{\odot}s^{-1} $ on to the black hole is considered, regarding the flux definition, the accumulated magnetic flux at $ r=R/R_{g}=3 $ would be of the order of $ \Phi_{30}\approx 0.02 $. Now, we assume that such a magnetic flux is accumulated during hyperaccretion and that it does not change with time. On the other hand, \citet{Yi et al.2016} have found that the $ 0.3-10 keV $ isotropic energy of X-ray flares is mainly distributed from $ 10^{50} $ to $ 10^{52} $ erg, which is more than two orders of magnitude less than that of the prompt emission of GRBs. Therefore, leading to $r_{m}=33$, the adoption of $ 10^{-3}~M_{\odot}/s^{-1} $ for the late time evolution mass supply rate sounds to be a reasonable choice. Another assumption that we made is the value of $10^{-3}$ for $\epsilon$. Indeed, this parameter is on less firm footing (for more discussion about this parameter and its physical importance see, e.g., \citet{Ronning et al.2016}; \citet{Narayan et al.2003}; and \citet{McKinney et al.2012}). 
 
It is the time to make our estimates for the duration time scales of the late time activities. With $ \Phi_{30}\approx 0.02 $, the accretion rate would be about $0.04~M_{\odot}s^{-1}$. Clearly, this accretion rate is around three orders of magnitude less than the rate $\dot{M}=10~M_{\odot}s^{-1} $. Thus our outcome is well consistent with observations studied by \citet{Yi et al.2016}. On the other hand, the disc mass for $r$ between 3 and 33 can be obtained about $0.25~M_{\odot}$. If this disc mass is a result of slow mass accumulation during the late evolutionary stage, then it will take about 250s to rebuild the disc for the mass supply rate of $ 10^{-3}~M_{\odot}/s^{-1} $ and 6s to accrete all this mass at the disc accretion rate of $ 0.04~M_{\odot}/s^{-1} $. The former lies between 100 to 1000s, the interval that has been inferred as the duration time of the flares by \citet{Yi et al.2016}. Yet, the latter might be an underestimated value of the flare duration, since we assumed a relatively high constant accretion rate, which is obviously against the time dependency of the light curves' behavior during the flare.
 
  Although, our model seems to be capable of predicting outcomes compatible with the observed duration time scale, there might be some considerations to which we had better point. Beside being affected by some other factors like rotation (\citet{Ronning et al.2016}), we are of the opinion that the magnetospheric radius should be influenced by the fragmentation. This process will lower the disc accretion rate which might result in a larger magnetospheric radius (inferable from $r_{m}$ relation) and, subsequently, in a longer duration time scale. Of course, such a correlation appears to agree with the observational evidences found by \citet{Ramirez-Ruiz et al.2001} and the analytical estimates studied by \citet{Ronning et al.2016}, in which the quiescent time-pulse duration correlation has been discussed. Finally, in agreement with our previous results (deduced from Figure~\ref{fig4}), the fragmentation process might lower the magnetic barrier effectiveness to produce X-ray flare.

\section{Conclusions}
\label{5}
                                                    We study the structure and evolution of neutrino dominated accretion discs, in which the consideration of self-gravity and magnetic field provide us with a more realistic picture of these central engines of GRBs. We find self-gravity a booster of magnetic field, mainly in the outer disc. Such an effect, especially in higher accretion rates, enhances the BP power and descends the neutrino luminosity fraction. The latter is against \citet{Liu et al.2014} outcomes but seems a natural result of the strong magnetic field presence as we discussed formerly. On the other hand, the probable fragmentation process, lessens the magnetic barrier possibility by a decrease in magnetic field diffusion time scale in higher mass accretion rates. This result can also be deduced from the estimations, we have already made, and their comparison with the observational evidences. 
                                                                                                          
                                                                                                          On the whole, we find both MHD and neutrino processes effective enough ($10^{50-54}erg/s$) to produce GRBs' spectrum. Of course, in the case of higher accretion rates, the drop in neutrino efficiency (Figure~\ref{fig7}), as a result of a growth in neutrino opacity, may confirm a decrease in neutrinos capability to transport the energy outside. In the context of late time X-ray flares, the magnetic barrier process would be more probable to power such extended emissions in low accretion rates, because fragmentation is less likely to happen. Yet, this might not be the case for higher accretion rates as fragmentation can overcome magnetic barrier to produce energetic X-ray flares. 
                                                                                                          
                                                                                                          Nevertheless, what is presented in this work might be a simplified model from the central engines of GRBs that should be improved by the consideration of all other physical aspects of these engines. For instance, the unsteady structure of NDAFs, resulting in jets with shells of different Lorentz factor, can realize our physical perception of NDAFs (as proposed by \citet{Narayan et al.1992}; \citet{Paczynski et al.1994}; \citet{Meszaros et al.1994}). The fact that viscous timescales in the inner disk, where all the neutrinos processes become important, are shorter than those in the outer disk, where $\dot{M}$ is expected to vary (inferable from Figure~\ref{fig4}), makes the steady approximation rather justified to be applied (\citet{Di Matteo et al.2002}), however, the time dependent behavior of the GRB engine beside the probable instabilities, cause the unsteady state to be a more viable approach. Or, considering the fully self-gravitating magnetized NDAF might enhance the validity of our estimations and results. On the other hand, several other mechanisms make efforts to describe X-ray flares such as fragmentation of a rapidly rotating core (\citet{King et al.2005}), differential rotation in a post-merger millisecond pulsar (\citet{Dai et al.2006}),  He-synthesis-driven wind (\citet{Lee et al.2009}), jet precession (\citet{Liu et al.2010b}), episodic jet produced by the magnetohydrodynamic mechanism from the accretion disc (\citet{Yuan et al.2012}), all of which might lead us to different outcomes.
                  
                                            One more point deserves a mention is that the high rotation, required to form the centrifugally supported disc that powers the GRB, should produce gravitational waves via bar (e.g., \citet{Dimmelmeier et al.2008}) or fragmentation instabilities that might develop in the collapsing core (see e.g., \citet{Ott2009}) and/or in the disc (\citet{Kobayashi et al.2003}; \citet{Piro et al.2007}). This gives our model the potential to affect the predictions and estimations made in the context of gravitational waves, that might be another complementary discussion to be worth noting for the future studies.  
                                                                                                                                                        
                                                                                                          Any way, considering such a variety of models or physical features, it is a matter of more information from the multi-band observations and the detections on the polarization and gravitational waves on the GRBs and their flares to decide which model or what physical conditions can provide us with a more accurate insight into the enigmatic nature of GRBs.

\acknowledgments

We are grateful to the constructive and thoughtful comments of the referee, which helped us to improve the early version of the article. This work was supported by the Ferdowsi University of Mashhad under the grant 43650 (1396/02/6).

\end{document}